# Adiponitrile-LiTFSI solution as alkylcarbonate free electrolyte for LTO/NMC Li-ion batteries

Douaa Farhat[a], Fouad Ghamouss[a] z, Julia Maibach[b], Kristina Edström[b] *, Daniel Lemordant[a] *

**Abstract:** Recently, dinitriles (NC(CH₂)ₙCN) and especially adiponitrile (ADN, n=4) have attracted the attention as secure electrolyte solvents due to their chemical stability, high boiling points, high flash points and low vapor pressure. The good solvating properties of ADN toward lithium salts and its high electrochemical stability (~ 6V vs. Li/Li⁺) make it suitable for safer Li-ions cells without performances loss. In this study, ADN is used as a single electrolyte solvent with lithium bis(trimethylsulfonyl)imide (LiTFSI). This electrolyte allows the use of aluminum collectors as almost no corrosion occurs at voltages up to 4.2 V. Physico-chemical properties of ADN-LiTFSI electrolyte such as salt dissolution, conductivity and viscosity were determined. The cycling performances of batteries using Li₄Ti₅O₁₂ (LTO) as anode and LiNi₁/₃Co₁/₃Mn₁/₃O₂ (NMC) as cathode were determined. The results indicate that LTO/NMC batteries exhibit excellent rate capabilities with a columbic efficiency close to 100%. As an example, cells were able to reach a capacity of 165 mAh.g⁻¹ at 0.1C and a capacity retention of more than 98% after 200 cycles at 0.5C. In addition, electrodes analyses by SEM, XPS and electrochemical impedance spectroscopy after cycling confirming minimal surface changes of the electrodes in the studied battery system.

## Introduction

Rechargeable Li-ion batteries have been widely developed in various commercial fields, and are excellent candidates for energy storage systems of electric vehicles and electronics applications [1][2][3]. In recent years, the need for high performances lithium-ion batteries has increased dramatically for use in electric vehicles (EVs) and hybrid electric vehicles (HEVs) [4][5]. Using cathodes working at high voltage (>4.5V vs. Li/Li⁺) is a way to increase the energy densities of batteries but alkyl carbonates commonly used as electrolyte solvents are not stable towards oxidation at these high potentials, hence limiting the battery's cycle life.
One of the most important challenges for further development of Li batteries for electric vehicles is safety, which requires the development of nonflammable electrolytes with no impact on the electrochemical window, cycle lifetime and stored energy. Considerable research efforts in Li-ion battery technologies have already been done in order to find alternative electrolyte solvents,
which are able to enhance safety and resolve issues related to the high flammability and reactivity of alkyl carbonates with flash points below 30°C [6]. In addition, the autocatalytic decomposition of LiPF₆, the most commonly used electrolyte salt, has to be addressed to remove safety hazards.
Aliphatic dinitriles (NC(CH₂)ₙCN) belong to a solvent family, which may be used as electrolyte solvent and replace, at least partially, conventional electrolytes based on alkyl carbonates. Recently, some reports have pointed out the advantages of dinitriles as electrolyte solvents owing to their high boiling point, high flash point and low vapor pressure [7][8][9]. In addition, their wide electrochemical window (~ 6V vs. Li/Li⁺) [7][8][10] makes them suitable for use with high voltage cathodes to increase the energy density. Dinitriles exhibit also good solvating properties due to CN groups and are commercially available at a relatively cheap price with a high degree of purity. Among aliphatic dinitriles, adiponitrile (ADN, n=4) has been already proposed as electrolyte solvent [7], with lithium bis-(trifluoromethanesulfonyl) imide (LiTFSI) as lithium salt in a MCMB (mesocarbon microbeads)/LiCoO₂ cell. However, because of the inability of ADN to form a stable and conductive solid electrolyte interphase (SEI) layer on the negative electrode, ethylene carbonate (EC) or SEI forming additives have been added to the electrolyte blend [11]. Therefore, in all previous studies ADN was not used as single electrolyte solvent but only with SEI-forming additives or co-solvent [7][12].

**Table 1.** Physico-chemical properties of adiponitrile (N≡C-(CH₂)ₙ-C≡N) (n=4).

| Sovent | MW(g.mol⁻¹) | εᵣ | μ (D) | η (mPa.s) | ρ |
|---|---|---|---|---|---|
| ADN | 108 | 30 | 4.12 | 6.3 | 0.97 |

Lithium Titanate (Li₄Ti₅O₁₂, LTO) has been investigated as one of the most convenient anode material for energy storage systems such as HEV [13][14]. LTO is an alternative anode to carbonaceous compounds [15] because of the flat discharge and charge plateaus at 1.55V vs. Li/Li⁺ and the fast transport of Li-ions throughout the spinel structure. The relatively high potential of this electrode increases the security as no dendritic lithium deposit is expected even at high power rates [16][17]. LTO belongs to the group of so-called ''zero-strain insertion material'' that show almost no change in lattice parameters (less than 0.2% in volume) during intercalation which makes it an ideal candidate for long-life lithium ion batteries [17][18]. According to Eq. 1 below, LTO is able to intercalate three lithium ions per formula unit with a theoretical capacity of 175 mAh.g⁻¹ [17][19] :

$$Li_4Ti_5O_{12} + 3Li^+ + 3e^- \rightarrow Li_7Ti_5O_{12} \quad (1)$$

Moreover, lithiation and delithiation of the LTO electrode occurs with an excellent reversibility and a columbic efficiency close to 100% since no significant reduction of the electrolyte nor the formation of a SEI layer is expected. Li-ion batteries using a high voltage spinel cathode material with an LTO anode are therefore considered as promising energy storage system with improved cycle life and rate capability as compared with graphite-based Li-ion batteries [20]. LiNi$_{1/3}$Co$_{1/3}$Mn$_{1/3}$O$_2$ (NMC) is today one of the most common large-scale commercial cathode materials for Li-ion batteries [21][22][23]. The layered structure of NMC can provide high specific capacity and fast kinetics [24] according to the following electrochemical reaction (2):

$$Li(NMC)O_2 = Li_{1/2}(NMC)O_2 + ½ Li^+ + ½ e^- \quad (2)$$

[a]    D. Farhat, Dr. F. Ghamouss, Prof. D. Lemordant
       PCM2E Laboratory, EA 6296, Department of Chemistry, UFR
       Sciences et Techniques, Université François Rabelais de Tours,
       Parc de Grandmont, 37200 TOURS, France
       ᶻ E-mail: ghamouss@univ-tours.fr
[b]    Dr. J. Maibach, Prof. K. Edstrom
       Department of Chemistry - Angström Laboratory, Uppsala
       University, Box 538, SE-75121 Uppsala, Sweden
       * Electrochemical Society Active Member
       Supporting information for this article is given via a link at the end of
       the document.

Because of its high voltage (3.7 V- 3.9 V) good cyclability, structural stability, safety and cost NMC is a good choice of cathode material when combined with LTO [21][25][26][27][28].

In this study, an ADN-LiTFSI solution has been used as an electrolyte in LTO/NMC batteries as such a configuration allows to combine the benefit of safety, high energy density (high voltage cathode), stable electrolyte (no solvent oxidation on NMC, no SEI formation on LTO) at a moderate cost. ADN-LiTFSI electrolyte physico-chemical studies such viscosity, ionic conductivity, thermal properties will be reported first. Then, the electrochemical performances of a LTO/NMC battery using this electrolyte will be discussed in terms of specific capacity, cell impedance, and cycling stability. Aluminum current collector corrosion in the present electrolyte is also examined by mean of chronoamperometry and mass loss during polarization at high voltage. Finally, electrodes are examined post-mortem using X-ray photoelectron spectroscopy (XPS) and scanning electron microscopy (SEM) in order to analyze the surface composition and morphology of the cycled electrodes.

## Results and Discussion

### 1. Characterization of the electrolyte

Figure 1 shows the evolution of the ionic conductivity at 25 °C with the salt concentration. The curve exhibits a flat conductivity maximum at a concentration of 0.75M to 1.25M. The maximal conductivity of 2.3 mS.cm$^{-1}$ is obtained around 1 mol.L$^{-1}$ in salt which is enough for practical use in Li-ion batteries. At higher salt concentration, the ionic conductivity of the electrolyte is limited by the increase in viscosity of the resulting solution and also by the formation of ions pairs and higher aggregates.

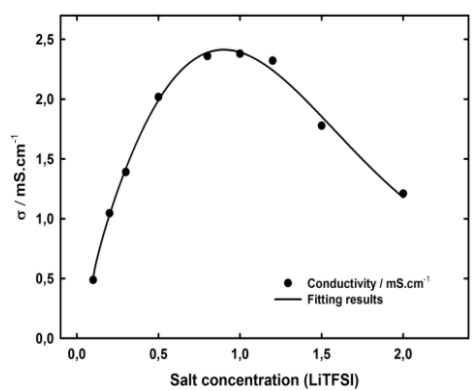

**Figure 1.** Conductivity of ADN electrolyte solutions vs. LiTFSI concentration.

The solubility of LiTFSI in ADN at room temperature is as high as 3 mol.L$^{-1}$. Solvation of lithium ions can be investigated by Raman spectroscopy as nitriles are characterized by the C≡N group stretching frequency, which occurs at 2260-2240 cm$^{-1}$ in aliphatic nitriles. In the case of pure ADN, the vibration is observed at 2243.75 cm$^{-1}$. When the C≡N is coordinated to a metal ion like Li$^+$ the R-C≡N→Li group has a C≡N stretching band which is shifted to a higher frequency (2276 cm$^{-1}$). As seen in Figure 2.a, the vibration of free C≡N bonds is slightly shifted by 3.5 to 8.75 cm$^{-1}$ in the presence of the salt. The ratio of the intensity of the coordinated C≡N band to free C≡N band increased linearly with the concentration of Li$^+$ in adiponitrile as indicated by the graph reported in Figure 2.b.

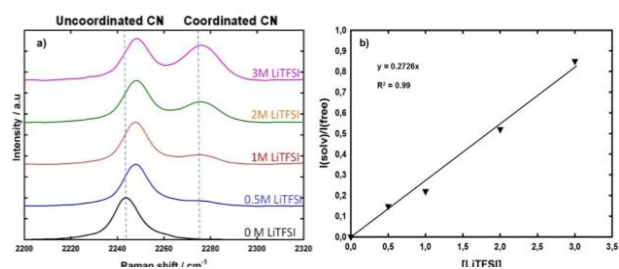

**Figure 2.** a) Raman spectra of LiTFSI solutions in ADN at room temperature, b) plot of the Raman intensity ratio of solvating to free CN groups I(solv)/Ifree as a function of the salt concentration.

From the slope of the regression fit and solution densities it is possible to determine the concentration (in mole per liter) of solvating and free CN groups as reported in Table 2. From these concentrations, it is also easy to calculate the mean solvation number of Li$^+$ by CN groups as a function of the salt concentration. Results reported in Table 2 show that at low concentration ([Li+]≤0.5mol.L$^{-1}$), the number of solvating CN group is 4 and the most probable structure for the solvated lithium is Li(NC-R-CN)$_4^+$, a monodentate complex. Nevertheless, at higher concentration in salt, the solvation number decreases owing to a lack of sufficient free solvating molecules and around [Li$^+$] = 2mol.L$^{-1}$, the number of solvating CN groups is only 2 indicating that the formation of ion pairs like (TFSI$^-$) Li(NC-R-CN)$_2^+$ becomes likely as simultaneously the conductivity drops sharply (refer to the following paragraphs). Even at 3 mol.L$^{-1}$, there are still free ADN molecules but the solvation number drops to 1.4.

**Table 2.** Concentration of coordinated and free CN groups as a function of LiTFSI concentration in ADN. [CN(solv)] : coordination number of CN groups per Li ion.

| [Li+] / mol.L$^{-1}$ | [CN(solv)]/ mol.L$^{-1}$ | [CN(free)]/ mol.L$^{-1}$ | [CN(solv)] per Li |
|---|---|---|---|
| 0 | 0 | 17.75 | 0 |
| 0.5 | 2.01 | 14.33 | 4.01 |
| 1 | 3.026 | 11.65 | 3.26 |
| 2 | 4.33 | 7.73 | 2.17 |
| 3 | 4.21 | 5.01 | 1.40 |

In the remainder of the study, the LiTFSI concentration has been fixed to 1 mol.L$^{-1}$ which is close to the maximum of ionic conductivity and also because such concentration is usually sufficient to avoid concentration polarization at moderate cycling rates [29].

Viscosity is an important factor to take into account for filling cells with electrolyte adequately (no bubbles) and rapidly. The temperature dependencies of the viscosity of ADN solutions containing LiTFSI together with their ionic conductivity are displayed in Figure 3.

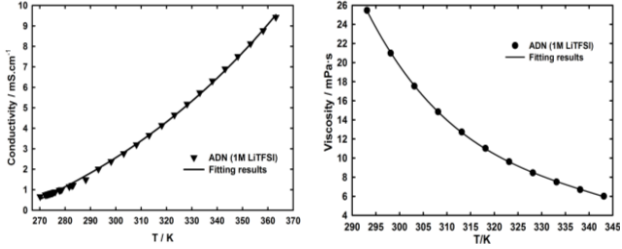

**Figure 3:** Evolution of conductivity (a) and viscosity (b) of ADN electrolyte solutions with 1M LiTFSI as a function of temperature (K). The line shows the fitting curves using equation 3.

The conductivity of ADN-LiTFSI (1M) is 0.77, 2.01 and 5.74 mS.cm$^{-1}$ respectively at 0 °C, 20 °C and 60°C and the viscosity of the same electrolyte decreases from 20.99 mPa.s at 25 °C to 7.50 mPa.s at 60 °C. As the Arrhenius equation does not provide a good fit for experimental data of both conductivity and viscosity, results are plotted using the Vogel-Tamman-Fulcher (VTF) equations respectively for conductivity (Eq.3) and viscosity (Eq.4):

$$\sigma = \sigma_0 \exp\left[\frac{-B_\sigma}{T-T_0}\right] \quad (3) \qquad \eta = \eta_0 \exp\left[\frac{B_\eta}{T-T_0}\right] \quad (4)$$

In Eq. (3) and (4), $\sigma_0$, $\eta_0$, $B_\eta$, $B_\sigma$, are fitting parameters and $T_0$ is the ideal glass transition temperature. The best fitting values for these parameters are reported in the Table 3, together with the correlation coefficient.

**Table 3.** Viscosity and conductivity VTF equation values for ADN-LiTFSI electrolyte

| | Viscosity | | | | Conductivity | | | |
|---|---|---|---|---|---|---|---|---|
| $\eta_0$ (mPa.s) | $B\eta$ (J.mol$^{-1}$) | $T_0$ (K) | $R^2$ | | $\sigma_0$ (mS.cm$^{-1}$) | $B\sigma$ (J.mole$^{-1}$) | $T_0$ (K) | $R^2$ |
| 0.141 | 675.75 | 163 | 0.9999 | | 117.792 | 453.67 | 183 | 0.9994 |

$R^2$ values reported in Table 3 show that the VTF fitting is convenient for this electrolyte and that the ideal glass transition temperatures are relatively close together for viscosity and conductivity processes taking into account an uncertainty of ±10K on $T_0$ values.

Electrolyte density is an important characteristic of electrolytes for practical reasons and especially at the industrial scale. Electrolyte densities were measured at atmospheric pressure as a function of the salt concentration and the temperature from 10 °C to 70 °C. Experimental results show the density ($\rho$) is, as expected, a linear function of the temperature T for all solutions as expressed in Eq. (5):

$$\rho(T) = a.T + b \quad (5)$$

The values of the parameters a and b in Eq. (5) are reported in Table 4.

**Table 4.** Density of LiTFSI solutions in ADN as a function of the temperature and salt concentration fitted par the equation: $\rho(T) = a.T + b$

| C / mol.L$^{-1}$ | a / g.cm$^3$.K$^{-1}$ | b / g.cm$^3$ |
|---|---|---|
| 0 | -7.10$^{-4}$ | 1.1782 |
| 0.5 | -8.10$^{-4}$ | 1.2531 |
| 1.0 | -8.10$^{-4}$ | 1.3093 |
| 1.5 | -8.10$^{-4}$ | 1.3983 |

The electrolyte densities increase with the salt concentration but decrease as expected with the temperature. Furthermore, as shown in the supporting information SI1, there is a linear relationship between the solution density and the molar concentration. Owing to this linear relationship, it is possible to estimate the density, $\rho$, over the solubility range of LiTFSI in ADN, i.e. from 0 to more than 3 mol.L$^{-1}$. At 25°C and 70°C, the relations between density and salt concentration are given respectively by equations (6) and (7):

$$\rho(25°C) = 0.1322 \, C_{LiTFSI} + 0.957 \quad (6)$$
$$\rho(70°C) = 0.1300 \, C_{LiTFSI} + 0.924 \quad (7)$$

Electrolyte density is an important parameter for batteries manufacturers for composition and purity control. The density of each cell component like active material, electrolyte and separator is also important for calculating batteries energy and power densities. $\rho$ is also a parameter in the Washburn equation [29] which is applied for determining the wettability of porous solid (separators and composite electrodes).

The thermal behavior of the electrolyte ADN-LiTFSI (1M) was investigated using DSC measurements in the range -60 °C to 20 °C. The DSC thermograms reported in Figure 4.a show that during the cooling cycle (20 °C to -60 °C), solidification of the supercooled liquids occurs abruptly at $T_c$= -20 °C for pure ADN and $T_c$= -30 °C for the LiTFSI electrolyte solution. Both peaks are exothermic and split in two close peaks. During the following heating cycle, pure ADN exhibits two successive melting peaks ($T_{m,1}$ and $T_{m,2}$) which may correspond to the two allotropic phases formed at the first cooling scan. In the case of LiTFSI solution in ADN, a glass transition temperature can be detected at $T_g$= -37 °C followed closely by a cold crystallization and the eutectic fusion temperature at $T_E$=-30 °C. The end of melting occurs at -5 °C on the liquidus line.

The thermal stability of the ADN and the ADN-LiTFSI blend has also been examined by TGA analysis (Figure 4.b). TGA results show that the solvent and its blend with LiTFSI undergo thermal degradation beginning at 178 °C and with a total mass loss of 99.0 % at 240 °C for ADN and 440 °C for the blend which undergoes a multi-step decomposition. The observed mass loss before 100 °C is attributed to ambient moisture and remaining water in the lithium salt.

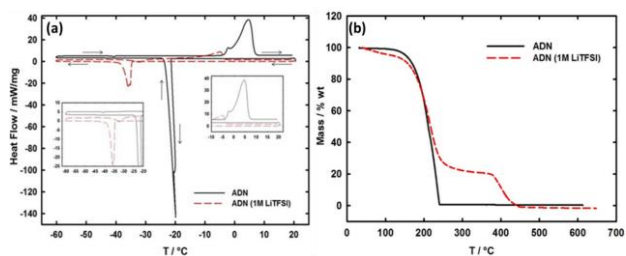

**Figure 4.** DSC (a) and ATG (b) curves for pure ADN and ADN with 1mol.L$^{-1}$ LiTFSI.

Aluminum, which is used almost exclusively as cathode current collector in rechargeable Li-ion batteries, is generally corroded in the presence of TFSI anions [30]. Indeed, Al dissolution has been observed when LiTFSI was dissolved in alkylcarbonates solvents. It has been shown that during its anodic polarization, Al is covered by a film which passivates its surface in LiPF$_6$ based electrolytes. This surface film is composed of AlF$_3$, Al$_2$O$_3$, and Al(OH)$_3$ which prevent aluminum from further corrosion. Using cyclic voltammetry, Yaser Abu-Lebdeh and al. [7], showed that [ADN] [LiTFSI] leads to the oxidation of Al but the oxidation current is decreased after a some CV cycles. Furthermore, the addition of EC to the electrolyte in order to ensure better battery performances has as consequence that aluminum corrosion is accentuated. Herein, we use chronoamperometry to study aluminum corrosion in ADN-LITFSI electrolyte. The chronoamperograms (current vs. time) obtained under a polarization of 4.2 V and 4.6 V vs. Li/Li$^+$ are presented in Figure 5*Figure 5*. Results obtained using Ethylene carbonate (EC)/ Dimethyl carbonate (DMC) with LiTFSI are also given for comparison. As seen in Figure 5, a huge oxidation current arises when the Al electrode is polarized at 4.2 V or more when EC/DMC-LiTFSI is used. This current is the consequence of Al dissolution. In ADN-LITFSI, the oxidation current is almost negligible at 4.2 V and remains low (less than 1 mA/cm$^2$) at 4.6 V.

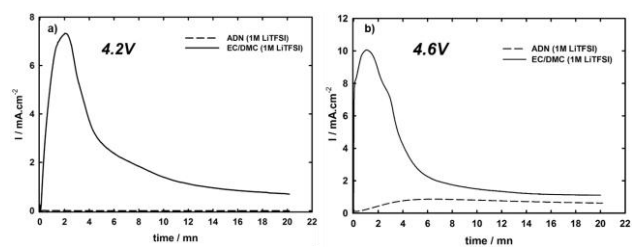

**Figure 5.** Current vs time of aluminum electrodes polarized at 4.2V (a) and 4.6V (b) in ADN-LITFSI and EC/DMC (1/1)-LiTFSI

At the end of the experiment (20 min), the electrode mass change has been determined using an SE2 Ultra-micro balance (Sartorius) as reported in Table 5.

**Table 5.** mass loss in percentage of the pristine electrode mass during polarization at 4.2V and 4.6V

|  | AND-LITFSI (1M) | EC/DMC-LiTFSI (1M) |
| --- | --- | --- |
| Polarization : 4.2V | ~ 0.0 % | 2.02 % |
| Polarization : 4.6V | 0.6 % | 3.6% |

Results in Table 5 show that at a polarization voltage of 4.2 V vs. Li/Li$^+$, the mass change, related to Al corrosion is negligible. In the following, electrochemical cycling tests of LTO/NMC cells are performed at voltages below this threshold potential, thus aluminum collector corrosion is not expected to contribute significantly to the electrochemical signal.

2. Electrochemistry of LTO/NMC full cells

Cyclic voltammograms were carried out at 25 °C in on a LTO/NMC cell containing ADN-LITFSI as electrolyte at scan rates ranging from 0.05 to 0.20 mV.s$^{-1}$. The voltage limits were set at 1.55 V and 2.80 V. As seen on the curves reported in SI2, the mean cell potential is 2.2 V and the cathodic and anodic peak intensities increase with the scan rate. Therefore, the peak current is controlled by diffusion and its value is given by the Randles–Sevcik equation [31][32][33]:

$$Ip = 2.69{\times}10^5 \; n^{3/2} \; A \; D_0^{\,1/2} \; V^{1/2} \; C \quad (8)$$

Where $Ip$ is the maximum peak current, $n$ the number of electrons exchanged (1 for LTO/NMC system), $A$ the surface area of the working electrode (cm$^2$), $D_0$ the diffusion coefficient (cm$^2$.s$^{-1}$), $v$ the scan rate (V.s$^{-1}$) and $C$ the bulk concentration (mol.cm$^{-3}$). The diffusion coefficient is calculated by plotting the value of the maximum current ($Ip$) intensity against the square root of the scanning rate. The linear correlation obtained indicates that the current is limited by diffusion in the solid phase (NMC cathode is limiting the current). The Li$^+$ diffusion coefficient is obtained from the slope of the graph: $D$ (Li$^+$) =3.19 10$^{-10}$ cm$^2$/s. This value is close to that obtained by others for NMC and other different electrolytes: 3.53 10$^{-10}$ cm$^2$/s in EC/DEC-LiPF$_6$ (1M)[34] and 3.502 10$^{-10}$ cm$^2$/s in EC/DEC/EMC-LiPF$_6$ (1M) [35].

It can also be seen that the cathodic and anodic peaks are shifted toward higher and lower values respectively, which means that the system is only quasi-reversible under these scan rates.

The first charge/discharge curves of LTO/NMC cells in ADN-LITFSI are displayed in Figure 6.a as a function of the applied charge and discharge rates (C/10 to C/2). The working potential limits are 1.6 V and 2.8 V. All cells exhibit good rate capabilities in both charge and discharge, but by increasing the current, the utilization of the active material decreases, and the capacities drop from 165 mAh.g$^{-1}$ at 0.1C (which is close to the theoretical values), to 145 mAh.g$^{-1}$ at 0.5C rate. Additionally, it can be seen that the coulombic efficiency of the sample is nearly 100%. Figure 6.b presents the first and the second galvanostatic charge/discharge cycles at C/20 rate in EC/DMC-LiTFSI for comparison. As a result, better cycling is achieved when LiTFSI is used as lithium salt in ADN than in EC/DMC. The huge irreversibility observed in this system is likely linked to side reactions, mainly current collector dissolution when LiTFSI is used alkyl carbonate solvents. This could lead to loss of contact between the active material and the current collector and then a decrease of the capacity of the battery.

Figure 7.a shows the discharge capacities of the LTO/NMC cells obtained at different charge/discharge rates (from C/20 to 1.5C). As shown from Figure 7.a, owing to the increase in the discharge current, the capacity of discharge declines from 170 mAh.g⁻¹ at C/20 to 80-100 mAh.g⁻¹ at 1C and even 60 mAh.g⁻¹ at 1.5C. Nevertheless, the initial capacity is recovered almost completely when the C-rate is set back to C/20.

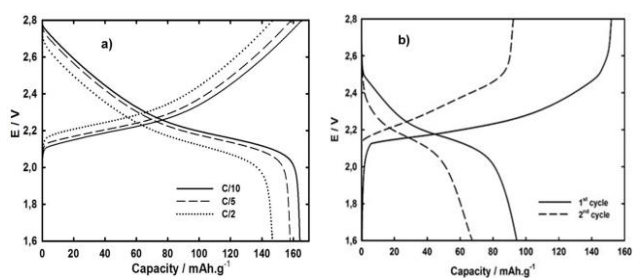

**Figure 6.** Galvanostatic charge discharge curves for LTO/NMC cell at different scan rates in (a) ADN- LiTFSI electrolyte, and (b) the first and the second galvanostatic charge discharge cycles at C/20 rate cycled in EC/DMC-LiTFSI.

The battery cycling performances have been also examined under a constant charging rate (C/10) and different discharging rates (C/10 to 2C). Results reported in Figure 7.b, show that in these conditions, the capacity retention of the battery is improved. The battery recovers more than 75 % of its initial capacity even at high discharge rate (2C).

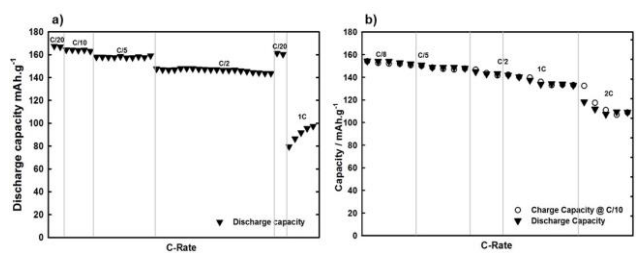

**Figure 7.** LTO/NMC cells cycled in ADN-LITFSI electrolyte: (a) Evolution of the discharge capacity at equal charge/discharge rates and (b) evolution of charge capacities (circle), discharge capacities (inverted triangle) when cells are charged at C/10 and discharged at C/8 to 2C.

Electrochemical impedance spectroscopy (EIS) has also been performed on the LTO/NMC cells and results are presented in Figure 8. EIS measurements were carried out over the frequency range 10 mHz – 1 MHz at the full charged state (2.8 V) after the first charge (left) and the second charge (right). The EIS spectra are composed of two partially overlapping semicircles at high to middle frequencies and a straight slopping line at low frequencies. The intercept of the first semi-circle with the X-axis represent the electrolyte resistance $R_s$. The two overlapping semi-circles at intermediate frequencies are linked to the electrode/electrolyte interfaces and to the charge transfer resistance $R_{ct}$ as well as to the related capacities. The straight line in low frequency range is attributed to the Warburg impedance W, which is related to lithium diffusion in the insertion electrodes. Both spectra are very similar from the first to the second charge which clearly shows that the interfaces are stable. However, the frequency of the second semi-circle changes from 37.3 Hz after one charge to 26 Hz after two charges. This is linked to a slight modification of the electrochemical double layer capacitance and/or charge transfer resistance.

**Figure 8.** Impedance spectra at the fully charge state 2.8V, after the first (left) and the second (right) charge for LTO/NMC cells cycled in ADN-LiTFSI electrolyte.

The EIS data have been fitted by the electrical equivalent circuit given in Figure SI3 in the supporting information.

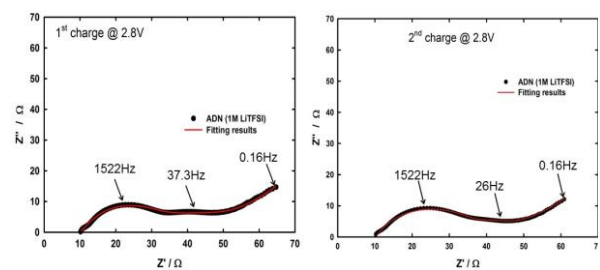

Extended cycling has been performed at the C/2 rate for charge and discharge over 200 cycles. The discharge capacities are plotted against the cycle number in Figure 9. ADN-LiTFSI electrolyte exhibits a fair cycling stability with a mean capacity of 135 mAh.g⁻¹ and a columbic efficiencies close to 100 %. Hence, this electrolyte can be used in LTO/NMC cells without almost no capacity fade for at least 200 cycles.

3.   Surface characterization:

The particle size and morphology of the active material is examined by scanning electron microscopy (SEM). In Figure 10, pictures of the LTO and NMC electrodes before and after cycling at 0.5C for charge and discharge are presented. SEM pictures of pristine electrodes show particles with flat surfaces and without any serious agglomeration. They display a homogeneous distribution in size ranging from 0.2 μm to 0.5 μm for $Li_4Ti_5O_{12}$ and 0.5 μm to 1.2 μm for NMC. After cycling, the same electrodes did not reveal any obvious cracks or other damage even after high C-rate testing. Nevertheless, a film is observed in the case of LTO electrode after being cycled in ADN-LiTFSI and this could be related to the shift in frequency of the second semi-circle observed by EIS.

In order to further investigate the surface layer formation indicated by EIS and SEM, the electrode surfaces of two LTO/NMC full cells using 1M LiTFSI in ADN electrolyte were characterized using hard X-ray photoelectron spectroscopy with an excitation energy of 2300 eV. The samples were collected after 1 charge from OCV to 2.8 V at C/10 (labeled "1 cycle") and after 2 cycles charge/discharge at C/10 (1.6 V - 2.8 V) followed by 10 cycles of charge/discharge at C/2 and a final charging step at C/2 (labeled "10 cycles"). Thus, in both cases the batteries were disassembled in charged state, leaving the LTO in a lithiated state and the NMC in a delithiated state.

The core level spectra relevant for determining whether or not surface film formation took place are shown in Figure 11.

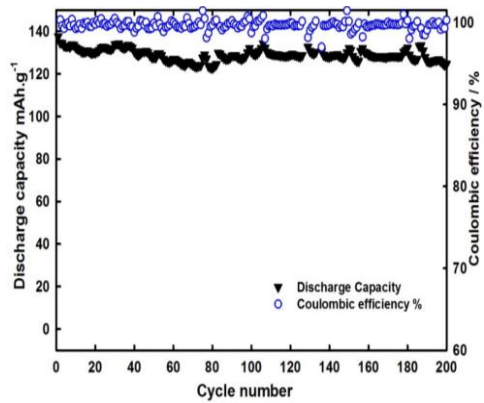

**Figure 9.** Evolution of the discharge capacity and the columbic efficiency of LTO /NMC cells at 0.5C charge/discharge rate, in ADN-LiTFSI electrolyte.

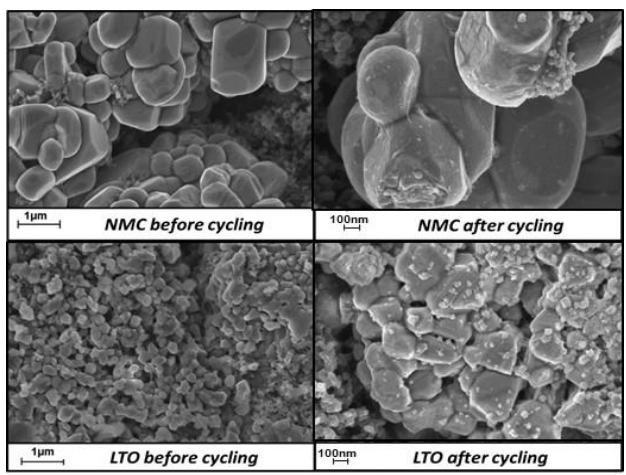

**Figure 10.** SEM images for NMC and LTO before cycling (left) and after being cycled in ADN- LiTFSI.

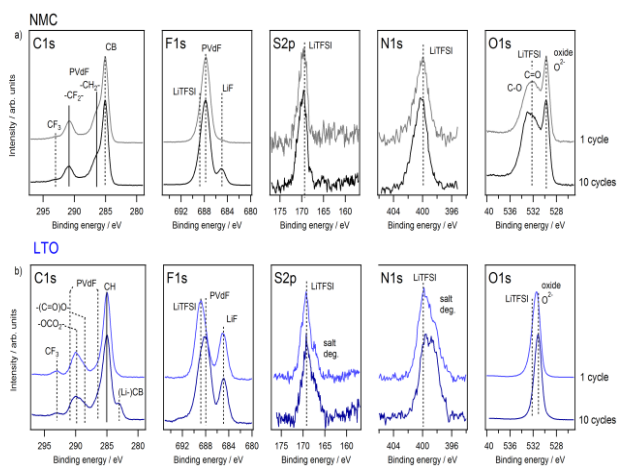

**Figure 11.** C1s, F1s, S2p, N1s and O1s spectra of NMC (a, top, in black) and LTO (b, bottom in blue) after 1 and 10 cycles. The spectra were recorded using 2300 eV excitation energy and are normalized in intensity to the maximum intensity.

The NMC spectra were calibrated in binding energy with respect to the $-CF_2-$ component of PVdF at 290.9 eV [36] as no clear adventitious hydrocarbon peak could be observed on these electrodes. The LTO samples on the other hand showed a very pronounced hydrocarbon peak (285 eV) which was used for binding energy referencing.

In general, the C1s spectra of the NMC electrodes show three main components: The two contributions of PVdF ($-CF_2-$ at 290.9 eV and $-CH_2-$ at 286.5 eV) and a strong carbon black signal at 285.05 eV. It is noteworthy, that the absolute binding energy of the carbon black peak is high compared to previous literature reports. However, this has been observed previously for $LiNi_{0.5}Mn_{1.5}O_4$ [37] in form of a reduced binding energy difference between PVdF and CB for cycled electrodes. Additionally, at 293 eV binding energy a low intensity peak from the LiTFSI salt ($-CF_3$) can be observed. The increased intensity between the two PVdF components is most likely due to C-O and C=O groups which are readily observed on the surfaces of CB powders. Overall, both carbon spectra after 1 and 10 cycles do not give a clear indication for a surface layer formation.

The situation is very different in case of the C1s spectra of LTO. The main emission after both 1 and 10 cycles is the hydrocarbon peak. Additionally, carbonyl (-(C=O)-O) and alkyl carbonate (-$OCO_2$-) peaks at 288.5 eV and 289.9 eV, respectively, contribute to the overall carbon intensity. Minor PVdF contributions are also observed as indicated. These features are generally associated with a SEI-type surface film formation. As the electrolyte solvent ADN does not contain alkylcarbonate (or other C-O) species, the oxygen species involved must stem from the surface termination of the CB powder or salt degradation products. Furthermore, the comparison between the LTO electrodes after 1 and 10 cycles shows, that the surface layer formation takes place already during the first charge but that its thickness might be reduced upon further cycling since after 10 cycles, a CB peak can be observed at 283 eV. The low binding energy of this carbon black peak could indicate partial lithiation of the carbon black. Another possibility could be that the bulk CB becomes visible after 10 cycles because of cracks in the surface layer. However, this was not observed in the SEM micrographs.

The F1s spectrum of the NMC samples is dominated by the binder signal at 688 eV. The peak asymmetry towards higher binding energies could be related to LiTFSI salt residues. After 10 cycles, a second peak is visible in the F1s region at 685 eV which is characteristic for LiF. As LiF is clearly visible on the LTO electrodes already after 1 cycle and since there are no further indications of surface layer formation on the NMC side, the LiF found on the NMC side could be the result of the migration of electrolyte decomposition products from the anode to the cathode during cycling. The most intense peak in the F1s spectra on the LTO side stems from LiTFSI and PVdF contributions. The change in maximum peak position between 1 cycle and 10 cycles towards lower binding energies could be due to the increased binder contribution due to a reduced surface layer thickness as was previously observed in the corresponding C1s spectrum.

In the S2p and N1s spectra from the NMC electrodes, single emissions corresponding to LiTFSI salt residues at 169.4 eV and 399.9 eV, respectively, can be observed. In case of the LTO electrodes, both the S2p and N1s salt peaks have shoulders on the low binding energy side indicating the deposition of salt decomposition products on the electrodes' surfaces. Thus a surface layer formation involving the electrolyte salt on the LTO side is further confirmed.

The O1s spectra are dominated by the metal oxide peak in both cases: The $O_{2^-}$ species in NMC leads to a peak at 529.8 eV while in LTO this peak is located at 531 eV. In the NMC case, the C-O compounds terminating the CB surface and possibly LiTFSI residues lead to a broad emission in the binding energy range of 532-535 eV. This, however, remains nearly unchanged further pointing towards the absence of a surface layer on NMC. In the LTO case, the oxide peak shows a much less pronounced asymmetry towards higher binding energies, which can be caused by salt residues and the previously observed carbonyl and alkyl carbonates.

In summary, both NMC and LTO show very comparable surface composition after 1 and 10 cycles as deduced from the PES spectra. In case of the NMC electrodes, no clear indications for surface layer formation could be observed. However, in case of LTO, the spectra indicate the formation of a surface layer involving electrolyte salt decomposition. Additionally, the surface layer on LTO seems to get reduced in thickness during cycling as the intensity of bulk characteristic CB increases with respect to the surface components.

## Conclusions

ADN-LiTFSI solutions have been proposed as safe electrolyte for Li-ion batteries. The optimal salt concentration is between 0.75M and 1.25M which corresponds to an ionic conductivity over 2 mS.cm$^{-1}$ at room temperature. The high solvating power of the CN groups of ADN toward Li ions permits to reach concentrated solutions, as high as 3M in LiTFSI. In these ADN-LiTFSI electrolytes, the number of Li$^+$ solvating molecules is reduced from 4 in diluted solutions to 1.4 near saturation (3M). Another advantage of the ADN-LiTFSI electrolyte is its excellent temperature stability from -30°C to almost 180°C.

ADN-LiTFSI electrolyte is well designed for use in batteries combining a stable LTO anode and a high potential NMC cathode. With this type of cell, it is possible to reach a specific cell capacity of 170 mAh.g$^{-1}$ at low rates (C/20) and even at high discharge rates (2C), the battery recovers more than 75 % of its initial capacity. The excellent stability of the electrolyte permits to perform at least 200 cycles with a coulombic efficiency close to 100% at each cycle. EIS spectroscopy as well as SEM and PES post-mortem experiments show little modifications of the electrode surface in good agreement with the high reversible capacity and high coulombic efficiency during cycling. Further work will consist in testing cells at high and low temperatures.

## Experimental Section

1.  Electrolyte, solvent mixture and cell preparation

Adiponitrile (ADN) (99%) was purchased from Aldrich, and used after being distilled under partial vacuum. Lithium bis-(trifluoromethanesulfonyl) imide (LiTFSI) is purchased from Solvionic (99.9% Extra Dry H$_2$O < 20ppm) and used as received. Electrolyte solutions were prepared in a glove box (MBraun) filled with argon containing less than 1ppm of water and oxygen.

Coated electrodes were provided from CEA (Grenoble, France, LMB Laboratory). LiNi$_{1/3}$Mn$_{1/3}$Co$_{1/3}$O$_2$ cathodes are composed of 90%wt. of active material (Umicore MX6), 6%wt. of Super P carbon and 6%wt. of PVdF5130 in organic medium, average loading ~14.1 mg$_{AM}$/cm$^2$. Li$_4$Ti$_5$O$_{12}$ anodes are composed of 90%wt. of active material, 5%wt. Super P carbon and 5%wt. PVdF5130 in organic medium, average loading ~13.6 mg$_{AM}$/cm$^2$. Current collectors for both electrodes are aluminum disks (1cm in diameter). Electrochemical measurements were carried out using coin-cells or two electrodes "Swagelok" systems. The electrodes are separated by a WHATMAN microporous glass-fiber paper (pore diameter 1.6 μm) filled with electrolyte. The electrolyte is obtained by dissolving LiTFSI at 1 mol.L$^{-1}$ in ADN in the glove box. Electrodes and separators were dried under vacuum for 24 hours at 80°C before assembling in the glove box.

2.  Experimental Methods

Differential scanning calorimetric (DSC) measurements were performed with a Perkin-Elmer DSC 4000 coupled with an Intracooler SP VLT 100. The instrument was calibrated with cyclohexane (solid–solid phase transition at −87.06 °C, melt transition (T$_m$) at 6.54 °C) and indium (T$_m$ at 156.60 °C). The measuring and reference cells are aluminum lids and pans. Each DSC thermogram was obtained under a N$_2$ atmosphere at a scan rate of 5 °C.min$^{-1}$ from 20 °C to −60 °C for cooling, and −60 °C to 20 °C for heating. The thermal gravimetric analysis (TGA) was carried out under nitrogen flux using a Simultaneous Thermal Analyzer STA6000 (Perkin Elmer). The samples were heated at a rate of 10°C.min$^{-1}$ from RT to 600°C. The electrolyte density was determined using an Anton Parr digital vibrating tube densitometer (model 60/602, Anton Parr, France). Viscosity measurements were performed using an Anton Parr rolling-ball viscosimeter (model Lovis 2000 M/ME, Anton Parr, France) using a 5 °C.min$^{-1}$ heating rate from 10 °C to 80 °C.

Ionic conductivity was measured at controlled temperature from -10 °C to 70 °C using a MCS 10 Fully Integrated Multichannel Conductimeter (BioLogic). Before measurements, the conductimeter was calibrated using KCl standard solutions (147 μS.cm$^{-1}$, 1413 μS.cm$^{-1}$, and 12.88 mS.cm$^{-1}$) at 25°C.

Raman spectroscopy measurements were collected using LabRAM HR Evolution High spectral resolution laser Raman spectrometer. The excitation source was a laser at 532 nm. The wavenumber of the Raman shift was calibrated by the spectrum of a standard Si wafer. The liquid samples were filled in high-quality glass capillaries having a 5 mm inside diameter and then tightly sealed in an Ar-filled glovebox to prevent any contamination from air. The measurements were carried out under a dry atmosphere at 25°C.

Cyclic voltammetry (CV) and galvanostatic charge-discharge were performed on a MPG2 multichannel potentiostat (BioLogic). CV measurements were operated at room temperature in the voltage range 1.6 V-2.7 V and at scan rate of 0.1 mVs$^{-1}$. Electrochemical impedance measurements were recorded on a multichannel potentiostat– galvanostat VMP system (BioLogic) at the end of charge and discharge of the batteries. The impedance was measured over the frequency range of 1 MHz to 10 mHz, with a signal amplitude of 5 mV.

The morphologies of the electrodes after cycling were investigated by Scanning electron microscopy using Zeiss ULTRA Plus Field Emission microscope.

To characterize the surfaces of the cycled electrodes, photoelectron spectroscopy (PES) in the tender X-ray regime (hν = 2300 eV) was performed. The measurements were carried out at the HIKE instrument (KMC-1 beamline) at the synchrotron facility BESSY II at Helmholtz-Zentrum Berlin. Prior to the measurements, the electrodes were retrieved from the batteries and rinsed with dimethyl carbonate (DMC) inside an argon filled glove box. The samples were then transferred directly from the glove box to the spectrometer in an air-tight transfer vessel to protect the sensitive surface from atmospheric conditions.

The research leading to these results has also received funding from the European Union's Seventh Framework Program (FP7/2007-2013) under grant agreement n° 608575 (Hi-C project).

We thank HZB for the allocation of synchrotron radiation beamtime.

**Keywords:** Adiponitrile, LiTFSI, Li-ion batteries, NMC, LTO.

## Acknowledgements

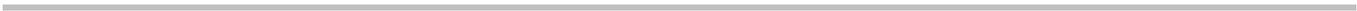

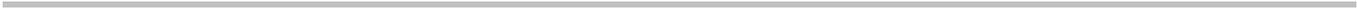